\documentclass[preprint]{aastex631}
\usepackage{amsmath,amsfonts,amssymb,amsthm}
\def\vec#1{\mbox{\boldmath $#1$}}
\begin{document}

\title{The Role of Small-Scale Flux Emergence in Pre-eruptive Magnetic Destabilization}

\author[0000-0001-5121-5122]{Satoshi Inoue}
\affiliation{Center of Solar-Terrestrial Research, New Jersey Institute of Technology \\
University Heights Newark, New Jersey 07102 USA}

\author{Huu Minh Triet Nguyen}
\affiliation{Center of Solar-Terrestrial Research, New Jersey Institute of Technology \\ 
University Heights Newark, New Jersey 07102 USA}

\begin{abstract}
Recent high-resolution observations have reported the emergence of small-scale magnetic flux shortly before the onset of solar flares; however, their role in solar eruption processes remains unclear. In this study, we perform three-dimensional magnetohydrodynamic (MHD) simulations to examine the effect of small-scale emerging flux on the eruption process.  We construct a bipolar active region by extrapolating a potential magnetic field from a pair of sunspots and drive the system with sustained rotational motions to build up strong shear along the polarity inversion line. A small-scale emerging flux, amounting to approximately $0.6\%-3.0\%$ of the total magnetic flux and having a polarity opposite to that of the main sunspot, is then introduced at the center of the bipole. As a result, although no significant differences are found in the pre-existing magnetic field during the early phase with or without the emerging flux, the current sheet formed in the sheared field lines collapses more rapidly, and the reconnection-driven eruption is initiated earlier in the case with small-scale emerging flux. This result suggests that small-scale emerging flux accelerates the destabilization of the pre-eruptive magnetic configuration through enhanced current-sheet thinning and magnetic reconnection, leading to an earlier eruption onset. Furthermore, we find that the involvement of highly twisted field lines during the pre-eruption phase plays a crucial role in determining the eruption onset time.
\end{abstract}


  \section{Introduction} \label{sec:intro}
Solar flares and coronal mass ejections (CMEs) are among the most violent phenomena in the solar system, during which large amounts of plasma and magnetic flux are ejected into interplanetary space (\citealt{Chen2011,Klipua2017}). When directed toward Earth, they can significantly disturb the geomagnetic environment and severely impact our technologically advanced society (\citealt{Temmer2021}). Therefore, understanding the fundamental physical processes behind these eruptions is crucial for improving space weather forecasting and mitigating their effects on human infrastructure.

Accurate prediction and early warning of solar flares and CMEs require a better understanding of their initiation mechanisms. Despite decades of research, the initiation processes remain an open question. It is widely believed that these eruptions are driven by the release of free magnetic energy stored in the solar corona (\citealt{Priest2002, Shibata2011, Toriumi2019}). Due to the high electrical conductivity of the coronal plasma, the magnetic field is effectively frozen into the plasma and becomes twisted or sheared as a result of photospheric motions (\citealt{vanBallegooijen1980, Amari_2003, Inoue2011}). As a consequence, sheared magnetic arcades or magnetic flux ropes (MFRs: \citealt{Filippov@2015AAJ,Chen2017,LiuR2020}), which is coherent structures of twisted magnetic field lines, are formed prior to the onset of solar eruptions (\citealt{Green2018, Patsourakos2020}). However, the precise physical mechanisms that drive the release of the stored magnetic energy remain poorly understood.

Newly emerging magnetic flux in vicinity of the active region prior to flaring has been observed and recognized as an important role in solar eruptions, including flares, filament eruptions, and CMEs \citep[e.g.,][]{Feynman1995,Wang&Sheeley1999,Tang2000,Li2000,Chen_2000,Dacie2018}. These studies suggest that reconnection-favored emerging flux appearing in the vicinity of a pre-existing filament or active region can destabilize the coronal field and initiate eruptions. \cite{LinJ2001} theoretically showed that the emergence of new magnetic flux can introduce additional equilibrium branches in coronal magnetic configurations and can lead to catastrophe-type eruptions depending on the strength and size of the emerging flux. More recently, \cite{ChenY2024,ChanY2025} extended this framework by coupling the catastrophe analysis to 2D resistive magnetohydrodynamic (MHD) simulations, reproducing one-step, two-step, and failed eruptions. 

MHD simulations have also been extensively performed to confirm the role of the emerging flux in the solar eruptions. \cite{Chen_2000} conducted a two-dimensional MHD simulation based on the observation results by \cite{Feynman1995} to find a role of the emerging flux in triggering process of the solar flares. The initial condition was set an equilibrium magnetic field in which the MFR is embedded in the potential coronal magnetic field.  They then gave ad hoc emerging flux at the polarity inversion line (PIL) on the solar surface under the pre-existing MFR or short distance away from the PIL. They found that the magnetic reconnection between the pre-existing magnetic field and emerging flux can alter the magnetic topology and break the equilibrium of the pre-existing magnetic field, leading to the eruption of the MFR. \cite{Torok2024} 
extended this model to three dimensions (3D) and investigated the eruption of a line-tied MFR whose two footpoints were anchored to the solar surface. They demonstrated that newly emerging magnetic flux can destabilize the pre-existing MFR by modifying the surrounding magnetic configuration and force balance, leading to its eruption. \cite{Kaneko@2014ApJ} demonstrated through two-dimensional (2D) MHD simulations that newly emerging magnetic flux can trigger an eruption through two different mechanisms, depending on its location relative to a pre-existing linear force-free arcade. When the emerging flux appears near the PIL, magnetic reconnection between the emerging flux and the overlying arcade can trigger an eruption. In contrast, when the emerging flux appears near the edge of the arcade, it compresses the pre-existing magnetic field, facilitating reconnection within the arcade and eventually leading to an eruption.

Recent space and ground-based high-resolution observations found the small-scale emerging flux just before flares (\citealt{Kubo2007, Wang_2017,Matsumoto2025b}). Since \cite{Bamba_2013} proposed that a complex magnetic structure formed in a localized region of the PIL can act as a trigger of solar flares, small-scale emerging flux introduced into a localized region of the PIL within an active region may also initiate flaring activity. \cite{Kusano2012} conducted a 3D MHD simulation started with sheared magnetic field lines. Their initial condition was designed as a linear-force-free field and surveyed the response of the force-free field with different small-scale emerging flux with different orientation. Consequently, they found two preferred emerging fluxes to cause the eruption.  These two emerging flux interact with pre-existing sheared field lines via magnetic reconnection, resulting into formation of highly twisted MFR and its subsequent eruption. Recently, this approach has been incorporated into data-constrained MHD simulations, which are based on the observed photospheric magnetic field. \citealt{Muhamad2017} applied small-scale emerging flux in a data-constrained MHD simulation to investigate flare triggering in active region (AR) 10930. Although their simulation successfully produced a highly twisted MFR following the scenario proposed by \citealt{Kusano2012}, it did not reach a full eruption. \citealt{Jing2021} also performed an data-constrained MHD for AR12665 simulation that included an emerging flux to investigate the role in the eruption; however, the small-scale emerging flux did not contribute to the eruption in their case.

Thus, a discrepancy still exists between the results obtained from idealized models and those based on observational data. Therefore, it is necessary to examine how effective small-scale emerging flux contributes to eruptions under more realistic magnetic configurations. In this study, we conducted MHD simulations of solar eruptions to investigate a role of small-scale emerging flux that extended 3D model by \cite{Kusano2012} a more realistically. In our model, we initialize an isolated bipole magnetic field at the solar surface and extrapolate the corresponding potential field. By imposing twisting motions on both polarities, the initially potential field is sheared and twisted, resulting in the formation of a sigmoidal structure (\citealt{Amari2000, Aulanier2010, Xia2014}) . The emerging flux is introduced at the PIL of the isolated bipole, above which sheared sigmoidal field lines are formed. We then examine the response of these sheared field lines to different magnitudes and radius of the emerging flux.

The rest of this paper is constructed as follows. The numerical method is described in section 2. The results and discussion are presented in section 3 and section 4. Finally, the conclusion are summarized in section 5.

\section{Numerical methods} \label{sec:num}
  \subsection{MHD simulation}
We numerically solved the zero-$\beta$ MHD equation, as described more in-depth by \cite{Inoue2014b} and \cite{Inoue2016}.

\begin{equation}
    \frac{\partial \rho}{\partial t}=-{\bf \nabla}\cdot(\rho {\bf v})+\xi{\bf \nabla^2\delta \rho},
\end{equation}

\begin{equation}
    \frac{\partial \vec{v}}{\partial t} = -({\vec v}\cdot{\vec \nabla}){\vec v} -\frac{1}{\rho} \vec{J} \times \vec{B} 
    + {\nu}\nabla^2 \vec{v},
\end{equation}

\begin{equation}
    \frac{\partial \vec{B}}{\partial t} = \vec{\nabla} \times (\vec{v} \times \vec{B}) + \eta \nabla^2 \vec{B} - \vec{\nabla} \phi,
\end{equation}

\begin{equation}
    \vec{J} = \vec{\nabla} \times \vec{B},
\end{equation}

\begin{equation}
    \frac{\partial \phi}{\partial t} + c^2_h \vec{\nabla} \cdot \vec{B} = - \frac{c^2_h}{c^2_p} \phi,
    \label{DednerEq}
\end{equation}

where $\vec{B}$ stands for the magnetic flux density, $\vec{v}$ for velocity, $\vec{J}$ for electric current density, and $\rho$ for plasma density where $\delta \rho$ is difference between $\rho$ and initial state $\rho_0$. 
The last equation was proposed by \cite{Dedner2002} and $\phi$ is the convenient potential used to remove the  numerical errors arising from $\vec{\nabla} \cdot \vec{B}$. The coefficients $\nu$ and $\eta$ are viscosity and resistivity respectively, set to $7.5\times 10^{-4}$ and  $1.0\times 10^{-5}$, and the coefficients $c^2_h$. $c^2_p$ in Equation (\ref{DednerEq}) are set to 0.04 and 0.01. The length, magnetic field, plasma density, velocity, and time are normalized by $L_0, B_0, \rho_0, V_A=B_0/\sqrt{\mu_0\rho_0}$, and $\tau=L_0/V_A$ respectively. 

Regarding the boundary conditions, the tangential components of the magnetic field, ${\bf B}_t$, evolve according to the induction equation, while the normal component, $B_n$, is fixed in time at each boundary. Specifically, the temporal evolution of $B_x$ and $B_y$ at the bottom is given by
\begin{equation}
\frac{\partial B_x}{\partial t}=\frac{\partial E_y}{\partial z}+\eta\frac{\partial^2 B_x}{\partial z^2}
-\frac{\partial \phi}{\partial x},
\label{bot_bx}
\end{equation}
\begin{equation}
\frac{\partial B_y}{\partial t}=-\frac{\partial E_x}{\partial z} +\eta\frac{\partial^2 B_y}{\partial z^2}
-\frac{\partial \phi}{\partial y},
\label{bot_by}
\end{equation}
 where ${\bf E}=-{\bf v}\times{\bf B}$. At each time step, the magnetic field at $k=1$, immediately above the bottom boundary $k=0$, is evolved according to the MHD equations. The updated magnetic field at $k=1$ is then used to evaluate the spatial derivatives in the induction equation at $k=0$, allowing us to evolve the horizontal magnetic field, $\vec{B}_t$($B_x$ and $B_y$), at the photospheric boundary according to equations (\ref{bot_bx}) and (\ref{bot_by}). The velocity is set to zero and the density is held constant, with the additional condition $\partial \phi/\partial n =0$ imposed at the boundaries. The prescribed magnetic field and velocity are applied during the twisting motions at the main polarities and during the emergence of the small-scale magnetic flux. Further details are described in the following section. 

The spatial derivatives are approximated using a second-order central difference scheme, and time integration is performed using the fourth-order Runge-Kutta-Gill method. A numerical box of $[1.0 \times 1.0 \times 1.0]$ size is given in its non-dimensional value and divided into $[420 \times 420 \times 420]$ grid points.

 \subsection{Initial condition $\&$ Twisting Motion}
\begin{figure}[h]
    \centering
    \includegraphics[width=0.9\textwidth]{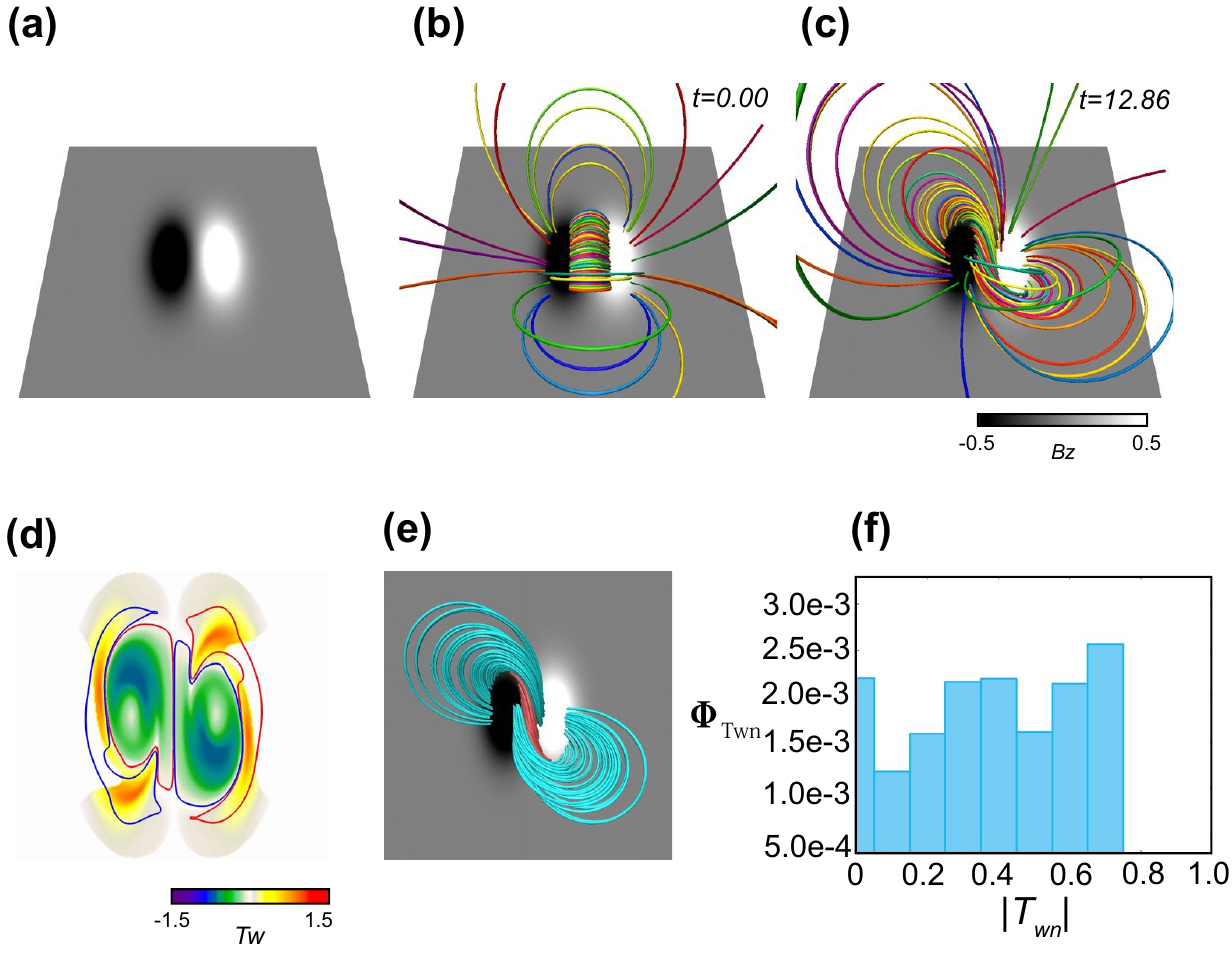}
    \caption{(a) $B_z$ component of the magnetic field on the surface, where white and black indicate the positive and negative polarities, respectively.(b) Potential field lines overlaid on the $B_z$ map, extrapolated from the surface $B_z$. (c) Twisted magnetic field lines formed by boundary twisting motions applied to an initially potential magnetic field, as shown in Figure \ref{f3}. These twisted field lines are given as the initial condition of the following MHD simulations. (d) Distribution of twist at the surface, where the red and positive indicate the positive and negative twist, respectively. The twist of each field lines is mapped on the photosphere. The red and blue lines represent contours of the vertical current density, corresponding to $J_z=1$ and $J_z=-1$, respectively. (e) The orange field lines represent shorter field lines with relatively small twist, whereas the light-blue field lines represent longer field lines with relatively larger twist.
    (f) This histogram shows the relationship between the magnetic flux contributed by regions with negative twist, $\Phi_{T_{wn}}$, and the magnitude of the negative twist component, $|T_{wn}|$.}
    \label{f1}
\end{figure}

To make an initial condition, having strong sheared field lines, we proceed following steps. First, we set up a simple bipole magnetic field sunspot in the center of the bottom boundary layer of the simulation box as shown in Figure \ref{f1}(a) and extrapolated the potential magnetic field (Figure \ref{f1}(b)) based on the Green function  method (\citealt{Sakurai1982}). 

Second, we solve the zero-beta MHD equations described above, applying twisting motions to the bottom boundary of the bipole field. At this stage, we adopt the following relation,
\begin{equation}
    \rho=|{\vec B}|,
\end{equation}
to evolve the density following \cite{Amari1996} and \cite{Inoue2014b}, which allows the twist to propagate rapidly in weak-field regions. This procedure is similar to that used in nonlinear force-free field extrapolations (\citealt{Inoue2016}). The initial velocity and density are given as $\vec{v}=0$, and $\rho =|\vec{B}_p|$, respectively, where $\vec{B}_p$ denotes the potential field, while $\phi=0$.

We imposed twisting motion in the bottom bipole field (\citealt{Amari2000, Aulanier2010, Xia2014}), which is  given as the following equation,
\begin{equation}
    \vec{v_h}^{(b)}|_{z=0}=(v_x,v_y)=\left(\frac{\partial \psi}{\partial y}, -\frac{\partial \psi}{\partial x}\right),
\end{equation}
where $\psi$ corresponds to the stream function that satisfies 
$\vec{v_h}^{(b)}|_{z=0}=\vec{z} \times \vec{\nabla}_h\psi$. $\psi$ is given as
\begin{equation}
    \psi(x,y,t) = \gamma(t) \left( B_z^{(b)}\right)^2 exp\biggl\{ \frac{-[(B_z^{(b)})^2-(B_{z_{max}}^{(b)})^2]}{(B_{z_{max}}^{(b)})^2} \biggr\},
\end{equation}
where $B_z^{(b)}$ and $B_{z_{max}}^{(b)}$ correspond to the normal magnetic field and its maximum value at the bottom surface. Following \cite{Aulanier2010}, $\gamma$ is given as follows, 
\begin{equation*}
   \gamma(t)=-0.5tanh \biggl\{2.0\frac{(t-t_{cri})-1.0}{0.5}\biggr\},
\end{equation*}
 here, $t_{cri}$  is the critical time at which the twisting motion is terminated and is set to $t_{cri}=15.5$ in this simulation. When the simulation time exceeds this value, $\gamma(t)$ gradually decreases to zero, and the twisting motion is eventually halted. The horizontal velocity ${\bf v}_h$ is prescribed along contours of $B_z$, ensuring that the normal magnetic flux, $\int\mathbf{B}\cdot d\mathbf{S}$ at the bottom boundary is conserved throughout the simulation. 

Figure \ref{f1}(c) shows the twisted field lines at $t=12.86$, formed by the twisting motion applied at the bottom boundary. The twisted field lines develop an inverse S-shaped sigmoidal structure, which is a result from negative helicity injection through the bottom surface. Sigmoidal magnetic morphology is frequently observed in X-ray images prior to solar flares ({\it e.g.}, \citealt{Kawabata2018}), along the polarity inversion line of the main polarities. 

Figure \ref{f1}(d) shows the distribution of the twist number $T_w$, where the twist number of each field line is mapped onto the bottom surface. 
The twist number is defined following \cite{Berger2006} as 
\begin{equation}
    T_w = \frac{1}{4\pi}\int\frac{\vec{\nabla}\times\vec{B}\cdot\vec{B}}{\vec{B}^2}dl,
\end{equation}
where $dl$ is a line element along the field line. The negative twist dominates in both polarities, which is consistent with the formation of an inverse S-shaped sigmoidal structure. In contrast, positive twist is distributed outside the region of dominant negative twist. The red and blue lines represent contours of the normal current density, $J_z$. The direction of the electric current is opposite in the negatively and positively twisted regions. These correspond to the direct and return currents, respectively (\citealt{Barczynski@2020}). The positive twist formed outside the dominant negatively twisted region is associated with the return current. Hereafter, we focus on the negatively twisted field lines formed by the dominant twisting motion at the bottom boundary and denote the negative twist as $T_{wn}$.

Figure \ref{f1}(e) shows the 3D magnetic field lines forming a sigmoidal structure in the negatively twisted region. The orange and blue field lines represent shorter, less twisted field lines and longer, more highly twisted field lines, respectively. We find that the shorter field lines with relatively small twist cross the central PIL, whereas the longer field lines with relatively larger twist are located on both sides of the shorter ones.  Figure~\ref{f1}(f) shows a histogram of the magnitude of the magnetic flux associated with $|T_{wn}|$. This result indicates that the maximum twist does not exceed 0.8, and that the sigmoidal structure is formed by moderately twisted field lines. We use this magnetic field as the initial condition for the MHD simulation, where the initial density is prescribed as $\rho_0=|\vec{B}|$, the velocity $\vec{v}$ is reset to zero  everywhere except at the bottom, where the twisting motion is imposed, and $\phi$ is inherited from the previous calculation.

\subsection{Model of Small-scale Emerging Flux}
\begin{figure}[h]
    \centering
    \includegraphics[width=0.9\textwidth]{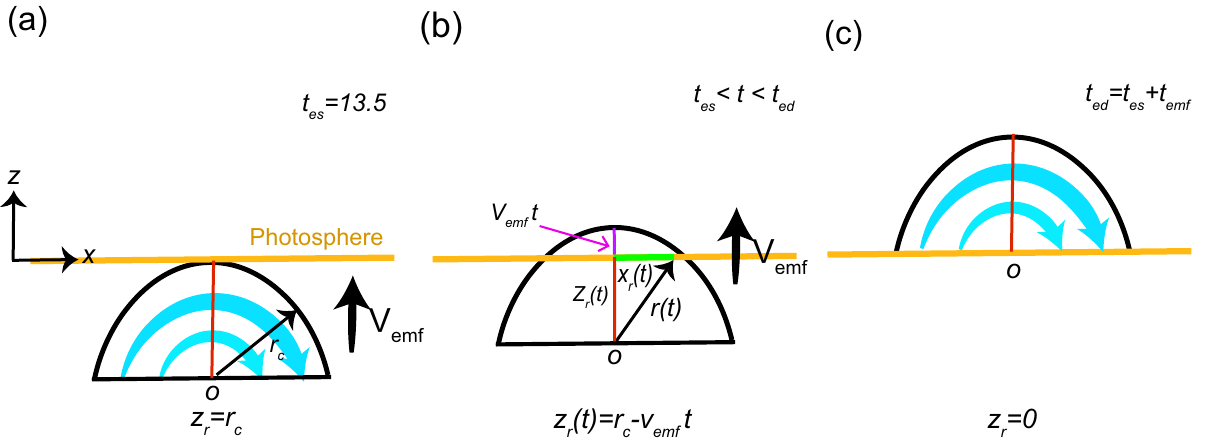}
    \caption{ Numerical model of the small-scale emerging flux.
     The small-scale emerging flux is modeled as a semi-circular structure with radius $r_c$, whose center is denoted by $o$. Blue arrows represent the magnetic field. (a) Initial state in which the small-scale emerging flux which is fully embedded beneath the bottom surface. The emergence starts at $t=13.5$. (b) Emergence of the small-scale magnetic flux through the solar surface. The emergence  velocity is given by $V_{emf}$. The distances 
     $z_r(t)$ and $r(t)$ denote the vertical distance from $o$ to the solar surface and an arbitrarily radial distance to the solar surface, respectively, while $x_r(t)=\sqrt{r(t)^2-z_r(t)^2}$. (c) Completion of the emergence of the small-scale emerging flux.} 
    \label{f2}
\end{figure}

Figure \ref{f2} shows an overview of small-scale emerging flux model. This follows previous model in \cite{Inoue2025}.
Small-scale emerging flux is introduced from the subsurface at $t=13.5$ at which the twisting motion still works on the bottom surface. As shown in Figure \ref{f2}(a), a semi-spherical magnetic structure (\citealt{Kusano2012, Muhamad2017, Jing2021, Inoue2025}), denoted as $B_e$, with radius $r_c$, was placed just beneath the photosphere with its center located at center of numerical box, $(x_e,y_e)=(0.5,0.5)$, and began to rise at a constant speed $V_{emf}$. An electric field $- {\bf V}_{emf} \times {\bf B}_e$ was imposed on the cross-section of the photosphere (bottom surface of the simulation box) to drive the semi-sphere upward at velocity ${\bf V}_{emf}$. The center of the emerging structure, denoted as point $o$, ascended from an initial depth of $z_r = r_c$ to $z_r=0$ at the photosphere, reaching the surface at time $\tau_{emf}=r_c/V_{emf}$.

\begin{figure}[h]
    \centering
    \includegraphics[width=0.75\textwidth]{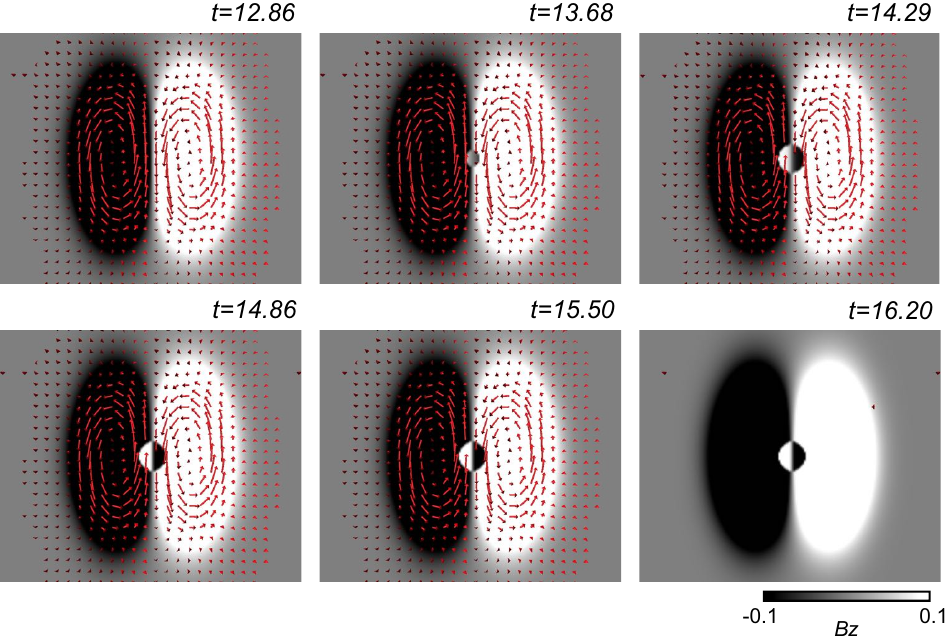}
    \caption{Evolution of sunspot motion and small-scale emerging flux in Run 1. White and black indicate positive and negative polarities, respectively, and red arrows represent the velocity. Even after $t=12.86$, which corresponds to the initial state of the MHD simulation, twisting motions are imposed at the surface, during which small-scale emerging flux is introduced locally in the main polarities. After the small-scale magnetic flux fully emerged, the twisting motions are halted. The polarity of the small-scale emerging flux is opposite to that of the main polarities.}
    \label{f3}
\end{figure}

The temporal evolution of the emerging flux is shown in Figures \ref{f2}(a)-(c), and the magnetic field components are given by:
\begin{equation}
B_{ex}({\bf r}, t) =  |B_{emf}|\frac{z_r(t)}{r(t)},
\end{equation}
\begin{equation}
B_{ey}({\bf r},t) = 0,
\end{equation}
\begin{equation}
B_{ez}({\bf r},t) = |B_{emf}|\frac{x_r(t)}{r(t)},
\end{equation}
where $|B_{emf}|$ is the magnitude of the emerging field, $z_r(t)$ is the vertical distance from the photosphere to the center $o$, and ($r(t)=\sqrt{x_r(t)~^2+z_r(t)^2}$) is the radial distance from point $o$ to a given position at the photosphere. In this study, the polarity of the small-scale emerging flux is prescribed to be opposite to that of the main polarity (see Figure \ref{f3}), consistent with the observations reported by \citealt{Wang_2017}.

The parameters used in this study are summarized in Table~1. Run R is the reference run, in which no small-scale emerging flux is imposed. In Runs 1–3, the magnetic field strength $|B_{emf}|$ and the radius ($r_c$) of the small-scale emerging flux are varied while keeping the total magnetic flux constant. In Runs 4–5, the magnetic field strength is increased while maintaining the same radius as in Run 1, corresponding to cases with larger total emerging flux. The emerging velocity $V_{emf}$ is set to $2.0\times 10^{-2}$ in all runs, and the corresponding emerging time $\tau_{emf}$ is listed in Table~1.
 
During the emergence, the surrounding magnetic field evolved under the boundary condition described above, while the magnetic field in the emergence region was fixed to the prescribed configuration. After the emergence completed, all bottom boundary fields, both within and outside the emerging region, evolved according to the boundary condition: the horizontal components evolved following the induction equation, while the normal component remained fixed in time. 

Figure~\ref{f3} shows the temporal evolution of the twisting motion and the small-scale emerging flux at the bottom boundary in Run 1. The twisting motion is already applied at the beginning of the MHD simulation ($t=12.86$). The small-scale emerging flux appears between $t=13.5$ and $t=14.5$. The twisting motion then gradually decreases starting at 
$t=15.5$ and is completely halted at $t=16.2$.

\begin{table}
  \begin{center}
  \caption{Parameters of the small-scale emerging flux for each Run. $r_p$, $r_c$, $|B_{emf}|$, $\Phi_{emf}/\Phi_t$, $V_{emf}$, and $\tau_{emf}$ denote the coefficient ranging from $0.0$ to $1.5$, the radius, the magnetic field strength, magnetic flux normalized by the total flux $\Phi_{t}$, the upward velocity of the emerging flux, and the emerging time, respectively. 
\label{tbl-1}}
  \begin{tabular}{crrrrrr}
  \tableline\tableline
   Simulation Runs & $r_p$ & $r_c$ & $|B_{emf}|$ & $\Phi_{emf}/\Phi_{T}$ & $V_{emf}$ & $\tau_{emf}$\\
  \tableline
   Run R    & $0.00$ & $0.00$ & $0.00$ & $0.00$ & $0.00$ & $0.00$ \\
   Run 1    & $1.00$ &$0.02r_p$ & $0.1/r_p^2$  & $0.006$ & $0.02$ & $1.00$  \\
   Run 2    & $1.25$ & $0.02r_p$ & $0.1/r_p^2$ & $0.006$ & $0.02$ & $1.25$ \\
   Run 3    & $1.50$ & $0.02r_p$ & $0.1/r_p^2$ & $0.006$ & $0.02$ & $1.50$ \\
   Run 4    & $1.00$ & $0.02r_p$ & $0.2/r_p^2$ & $0.012$ & $0.02$ & $1.00$ \\   
   Run 5   & $1.00$ & $0.02r_p$ & $0.5/r_p^2$ & $0.030$ & $0.02$  & $1.00$ \\     \tableline
   \end{tabular} 
   \end{center}
   \end{table}

  \section{Results} \label{sec:result}
\subsection{Overview of the Dynamics}
\begin{figure}[h]
    \centering
    \includegraphics[width=0.9\textwidth]{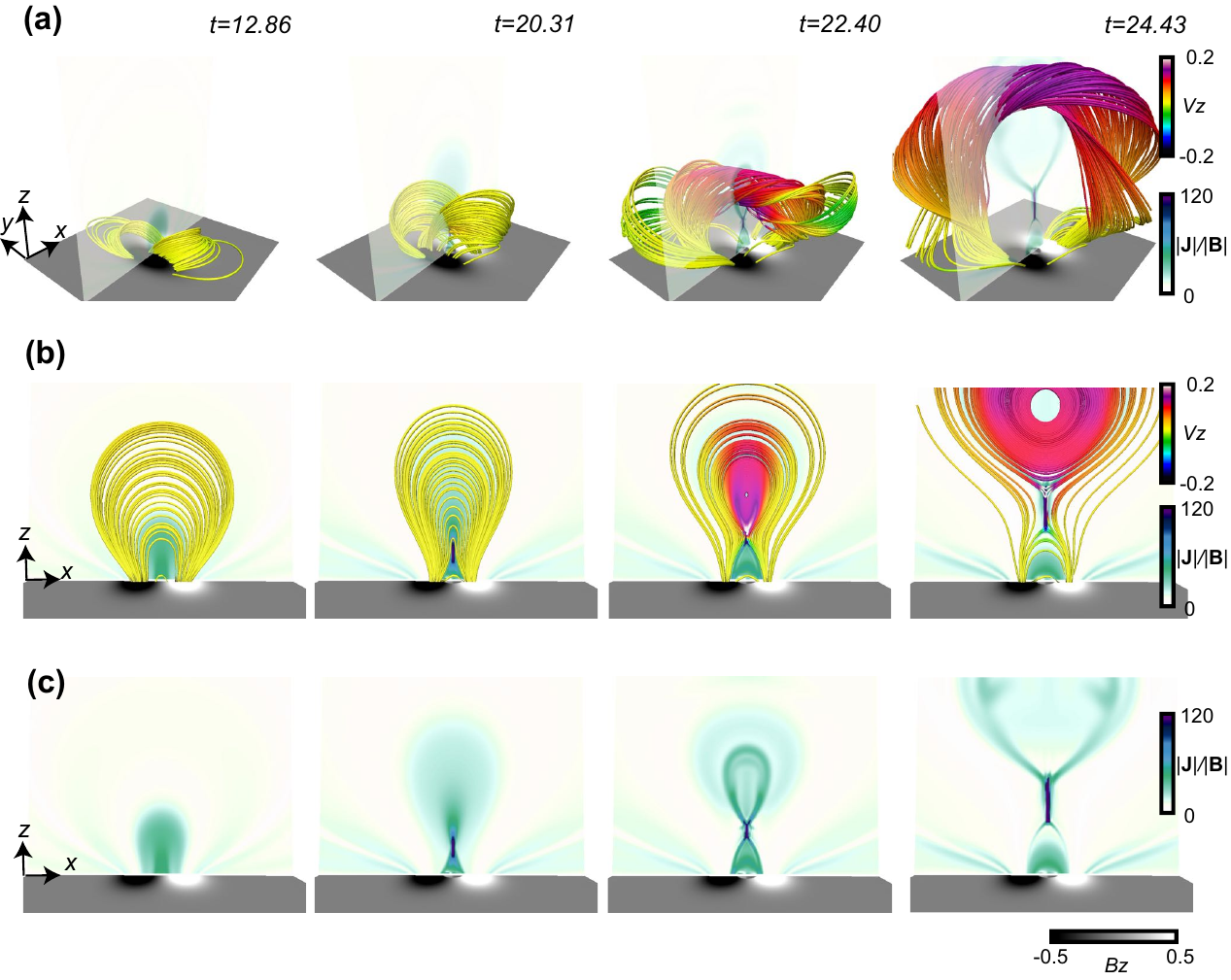}
    \caption{(a) Temporal evolution of three-dimensional magnetic field lines, with color indicating the vertical velocity $V_z$. The vertical cross section shows $|{\bf J}|/|{\bf B}|$, and the grayscale represents the $B_z$ distribution at the surface. (b) The field lines are projected onto the $x-z$ plane, and the vertical cross section shows $|{\bf J}|/|{\bf B}|$. The color coding is the same as in panel (a). (c) Temporal evolution of $|{\bf J}|/|{\bf B}|$, with same color coding as in panels (a) and (b).}
    \label{f4}
\end{figure}

Figure~\ref{f4}(a) shows the temporal evolution of magnetic field lines in Run~1, with the vertical velocity color-coded along the field lines. Figure~\ref{f4}(b) shows the projection of all field lines onto the $x–z$ plane, while Figure~\ref{f4}(c) displays the current density profile normalized by the local magnetic field strength. At $t=12.86$ which is start time of an MHD simulation, sheared field lines are formed along the PIL, and strong current density is concentrated there. After the twisting motion at the bottom boundary is halted, the field lines approach each other toward the PIL at $t=20.31$, forming strong, thin current layers. At $t=22.04$, magnetic reconnection leads to the formation of the MFR, corresponding to the O-point in Figure~\ref{f4}(b). The MFR then begins to erupt at $t=24.43$, during which a vertical current sheet is elongated beneath it.

\begin{figure}[h]
    \centering
    \includegraphics[width=0.9\textwidth]{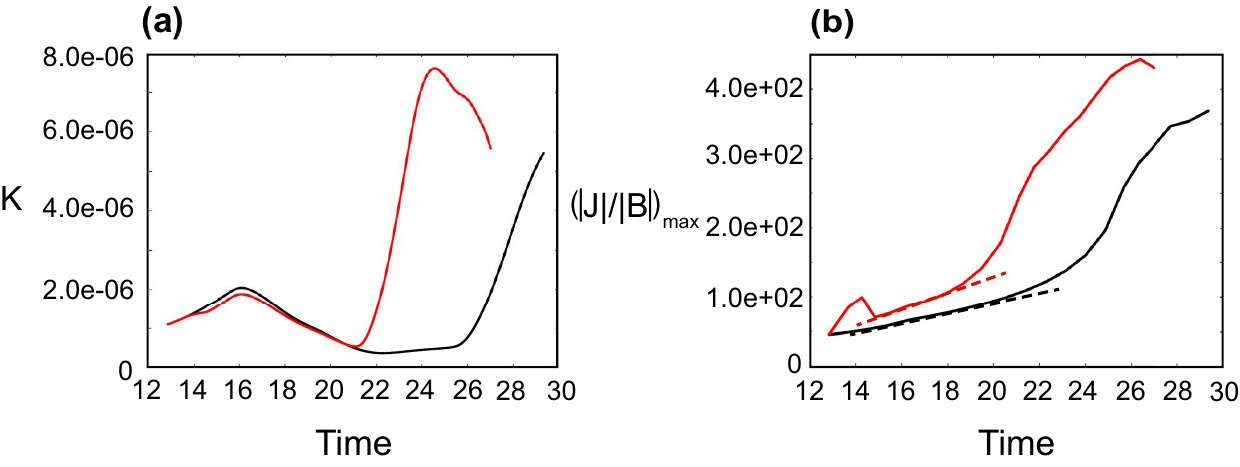}
    \caption{(a)-(b) Temporal evolution of the kinetic energy and the maximum value of $|{\bf J}|/|{\bf B}|$ at the center of the numerical box for Runs R and 1, shown in black and red, respectively. The black and red dashed lines serve as proxies for the growth of each current sheet for Runs R and 1, respectively, during the early phase of the evolution in each Run. }
    \label{f5}
\end{figure}

Figure~\ref{f5}(a) shows the temporal evolution of the kinetic energy integrated over the entire volume, $K=\int (\rho v^2/2)dV$, where $dV$ is the volume element. The black and red curves represent the results from Run~R and Run~1, respectively. Note that an eruption also occurs in Run R, even in the absence of small-scale emerging flux. This is likely due to numerical limitations, as the thin current sheet cannot be fully resolved at the available resolution. As shown in Figure \ref{f5}(a), the kinetic energy in Run 1 begins to increase rapidly around $t \sim 21$, whereas the corresponding increase in Run R occurs later, around $t \sim 26$. These results indicate that the onset of the eruptive phase occurs significantly earlier in the presence of small-scale emerging flux. These results indicate that small-scale emerging flux accelerates the transition from the pre-eruptive phase to the eruptive phase.

Figure \ref{f5}(b) shows the temporal evolution of maximum $|{\bf J}|/|{\bf B}|$ at the center of the numerical domain. The color coding is the same as in Figure~\ref{f5}(a). The temporary increase in Run~1 around $t=14$ arises from the interaction between the small-scale emerging flux and the pre-existing coronal magnetic field. Even in Run~R, which does not include small-scale emerging flux, the current sheet gradually intensifies, and once it reaches a certain level, an eruption is initiated. Although Run~1 follows a similar evolutionary trend, the development toward a thin and intense current sheet proceeds more rapidly than in Run~R. The dotted black and red lines, which serve as proxies for the growth rate of the current sheet, clearly show that the growth is faster in Run~1 than in Run~R. Therefore, small-scale emerging flux promotes the formation of a thin and intense current sheet and accelerates the destabilization of the pre-eruptive magnetic configuration. 

\subsection{Definition of Erupting Magnetic Flux Rope}
\begin{figure}[h]
    \centering
    \includegraphics[width=0.9\textwidth]{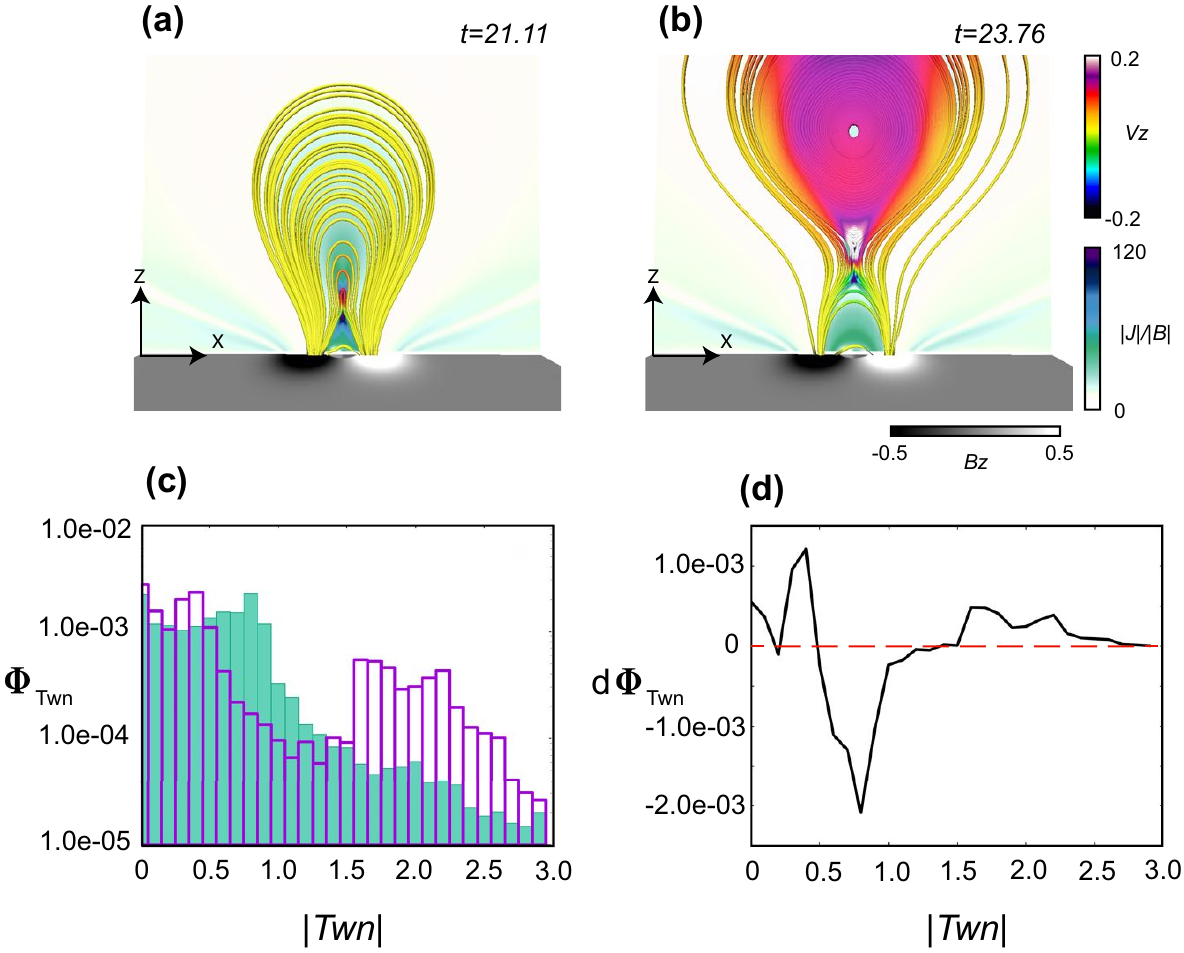}
    \caption{(a)-(b) Magnetic field lines projected onto x-z plane at $t=21.11$ and $t=23.75$, where the vertical cross section shows $|{\bf J}|/|{\bf B}|$. The color coding is the same as in Figure \ref{f4}. (c)This histogram shows the relationship between the magnetic flux contributed by regions occupied by negatively twisted field lines, denoted as $\Phi_{T_{wn}}$ and the magnitude of the negative twist, denoted as $|T_{wn}|$. The green and purple curves correspond to the results at $t=21.11$ and $t=23.75$, respectively. (d)
    Difference in the magnetic flux $\Phi_{T_{wn}}$ between at $t=21.11$ and $t=23.78$ as a function of the magnitude of $|T_{wn}|$. }
    \label{f6}
\end{figure}

Before we discuss the initiation of the eruption, we define the MFR used in this study, following \citealt{Inoue2025}. Figures~\ref{f6}(a) and (b) show the projected magnetic field lines before and after the formation of the MFR, whose center corresponds to the O-point of the magnetic field lines shown in Figure~\ref{f6}(b). Figure~\ref{f6}(c) presents a histogram of the magnitude of the magnetic flux associated with negatively twisted field lines as a function of the negative twist number, $|T_{wn}|$. We denote this magnetic flux by $\Phi_{T_{wn}}$. After the formation of the MFR, the magnetic flux associated with field lines having $|T_{wn}|$ in the range from 0.5 to 1.5 decreases, whereas that associated with field lines having 
$|T_{wn}| \ge 1.5$ increases. Figure~\ref{f6}(d) shows the difference in magnetic flux between $t=21.11$ and $t=23.76$. We denote the difference magnetic flux by $d\Phi_{T_{wn}}$. We find a transition around $|T_{wn}|=1.5$, where $d\Phi_{T_{wn}}$ changes sign from negative to positive. Since the erupting MFR with $|T_{wn}| \ge 1.5$ is formed through the reorganization of twisted field lines with $|T_{wn}| \le 1.5$, we define the MFR in this study as a cluster of twisted field lines with $|T_{wn}|\ge 1.5$.
 
\subsection{Initiation Mechanism: Instability vs. Reconnection}
\begin{figure}[h]
    \centering
    \includegraphics[width=0.9\textwidth]{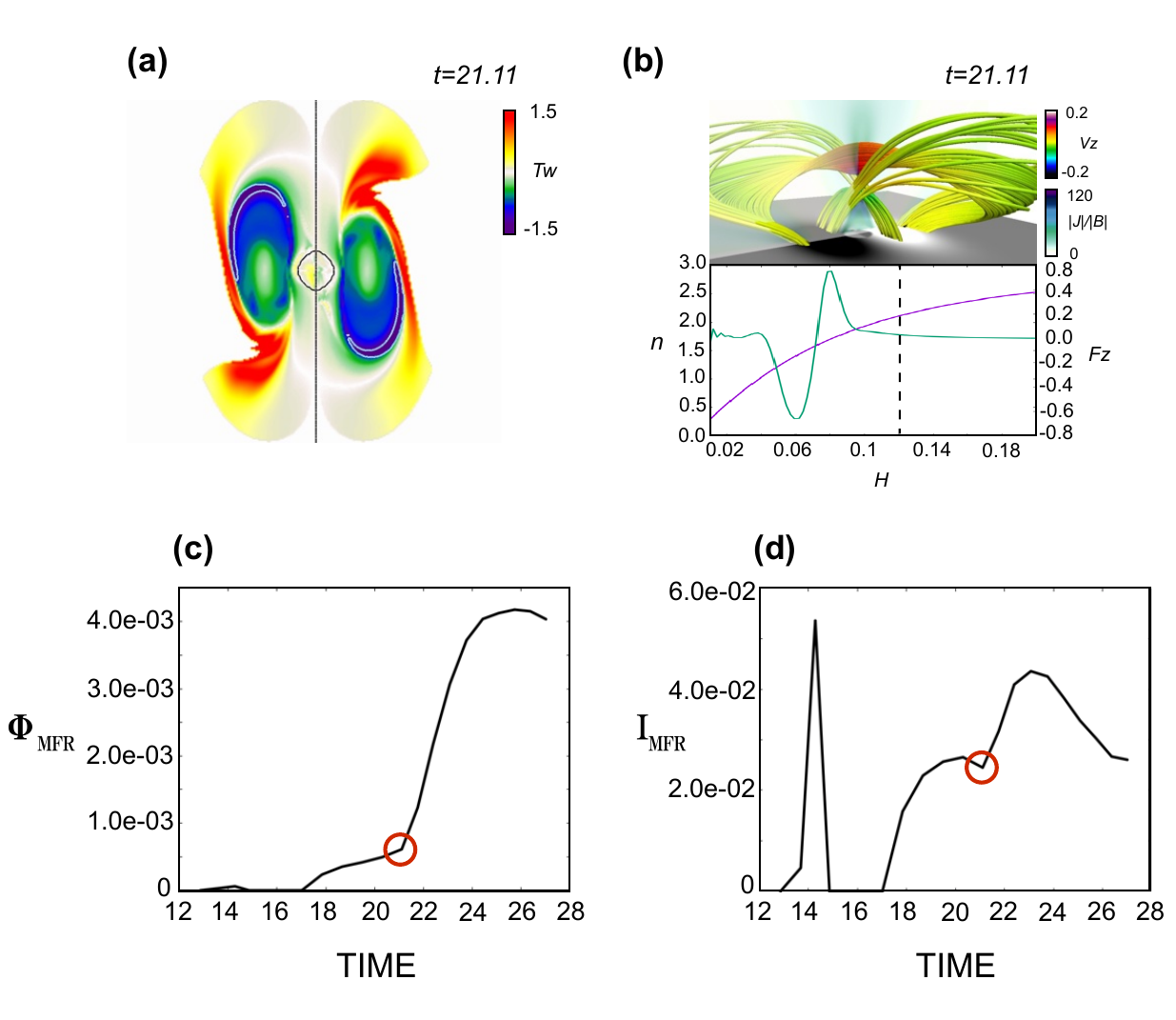}
    \caption{(a) Twist distribution where each field line is mapped on the bottom surface at $t=21.11$ in Run 1, which corresponds to just before the sudden eruption. White and black lines represent a contour of $T_w = -1.5$ and $B_z=0$, respectively. (b) The upper panel shows the 3D magnetic field lines and \(|{\bf J}|/|{\bf B}|\) on the vertical cross section at $t=21.11$ in Run 1, with the same color coding as in Figure \ref{f4}. This state corresponds to the stage immediately before the rapid eruption. The lower panel shows the height profiles of the decay index calculated from Equation (\ref{DI}) and the vertical component of the Lorentz force, $F_z$ in purple and green lines, respectively, evaluated along the vertical line passing through the center of the numerical domain. (c) Temporal evolution of the magnetic flux $\Phi_{MFR}=\int_{|T_{wn}|\ge1.5} |B_z| dS$ in Run 1, contributed by regions occupied by negatively twisted field lines with $|T_{wn}|\ge 1.5$. $dS$ is a surface element at the photospheric boundary. The red circle indicates time at which the magnetic field undergoes sudden eruption. (d) Temporal evolution of the integrated vertical current associated with the MFR at the photospheric boundary, $I_{MFR}=\int_{|T_{wn}|\ge1.5} |J_z| dS$ in Run 1, contributed by regions occupied by the negatively twisted  field lines with $|T_{wn}|\ge 1.5$. The red circle indicates time at which the eruption suddenly occurs.}   
    \label{f7}
\end{figure}

In this section, we clarify the initiation process by investigating the pre-eruption magnetic field in detail. Figure~\ref{f7}(a) shows the twist distribution at $t=21.11$ in Run 1, prior to the formation of the erupting MFR, where the twist of each magnetic field line is mapped onto the bottom surface. The white contours indicate $|T_{wn}|=1.5$, demonstrating that highly twisted field lines are already present before the eruption. The three-dimensional magnetic field structure is shown in the upper panel of Figure~\ref{f7}(b), where long, twisted field lines are formed above the current sheet and exhibit upward velocities, while sheared field lines sandwich the current sheet, facilitating magnetic reconnection.
 
We calculate the decay index to assess the stability of the MFR against the torus instability (\citealt{Kliem_2006}). 
Recent studies (\citealt{Myers2016, Zhong2021, Guo_L.H.2026a, Guo_J.H.2026b}) have pointed out that the external magnetic-field component parallel to the MFR axis, i.e., the toroidal component, can contribute to the confinement of the MFR and affect its eruptive evolution. Therefore, to evaluate the decay index relevant to the torus instability, we use the external poloidal magnetic field, defined as the component of the external field perpendicular to both the MFR axis and the eruption direction. In this study, we first determine the unit vector along the MFR axis, \(\mathbf{e}_{\rm tor}\), at the center of the MFR from the simulated magnetic field. Since the MFR erupts approximately in the vertical direction, the unit vector in the poloidal direction is defined as
\begin{equation*}
\bf{e}_{pol}=\frac{{\bf e}_z \times {\bf e}_{tor}}{|{\bf e}_z \times {\bf e}_{tor}|},
\end{equation*}
where \(\mathbf{e}_z\) is the unit vector in the vertical direction. The external poloidal magnetic field is then obtained from the potential magnetic field ($\vec{B}_{pot}$) as
\begin{equation*}
{\bf B}_{pol}=(\bf{B}_{pot}\cdot {\bf e}_{pol})\bf{e}_{pol},
\end{equation*}
and its magnitude,
\begin{equation*}
    B_{pol}= |\bf{B}_{pot}\cdot {\bf e}_{pol}|
\end{equation*}
is used to calculate the decay index. The decay index is defined 
\begin{equation}
    n = -\frac{z}{B_{pol}} \frac{\partial B_{pol}}{\partial z},
    \label{DI}
\end{equation}
which serves as an indicator of the poloidal field providing the strapping force. The decay index quantifies how rapidly the external poloidal component of the strapping field decreases with height. It is well known that an MFR becomes unstable to the torus instability when the decay index exceeds a critical value, typically $n \sim 1.5$ (\citealt{Kliem_2006, Aulanier2010, Fan_2010}).

 The lower panel of Figure \ref{f7}(b) shows the height profiles of the decay index calculated using Equation (\ref{DI}) and the vertical component of the Lorentz force, ($F_z$), both evaluated along the vertical line passing through the center of the numerical domain. At the height of the MFR axis, the vertical Lorentz force is positive but close to zero, suggesting that the MFR is approximately in force balance at this stage. Furthermore, the MFR axis is located above the height at which the decay index reaches the critical value of 1.5. These results indicate that the pre-eruptive MFR is in a configuration consistent with the onset condition for the torus instability. Therefore, the torus instability is a possible mechanism contributing to the subsequent eruption.  On the other hand, magnetic reconnection also appears to play a significant role in the eruption process. This raises the key question of which mechanism, torus instability or magnetic reconnection, acts as the primary driver of the eruption.
 
 To address this question, we examine the temporal evolution of the toroidal current, $I_{tor}$, flowing inside the MFR. Assuming a semi-toroidal geometry, the inductance and the toroidal current are denoted by $L$ and $I_{tor}$
 , respectively. The poloidal magnetic flux threading the semi-torus is then expressed as \begin{equation}
     \Psi = LI_{tor}.
     \label{flux}
 \end{equation}
The external inductance of a semi-torus can be written as a function of the major and minor radii, $R$ and $a$, as
\begin{equation}
  L(R,a) = \frac{\mu_0 R}{2} \biggl[ ln\left(\frac{8R}{a}\right)-2+\frac{l_i}{2} \biggr], 
\end{equation}
where $l_i$	is the internal inductance parameter. Under the condition $R>>a$, the inductance satisfies  $\partial L(R,a)/\partial R >0$, implying that the inductance increases as the major radius $R$ increases during the eruption. Within the framework of ideal MHD, the magnetic flux given by Equation~(\ref{flux}) is conserved, and therefore the toroidal current $I_{tor}$ must decrease as the eruption proceeds. In contrast, magnetic reconnection generates poloidal magnetic flux within the MFR (\citealt{Inoue2025}), which can suppress or increase the toroidal current within the MFR. Therefore, investigating the evolution of $I_{tor}$ is essential for addressing the question posed above.

First, we investigate the evolution of the magnetic flux associated with the highly twisted field lines comprising the MFR, which is evaluated by integrating $B_z$ over the region where $|T_{wn}|\ge 1.5$. Figure \ref{f7}(c) shows the temporal evolution of the magnetic flux of the MFR, 
\begin{equation}
\Phi_{|T_{wn}|\ge1.5}=\Phi_{MFR}=\int_{|T_{wn}|\ge1.5}|B_z| dS
\end{equation}
where $dS$ is the surface element on the bottom surface. This magnetic flux is associated with highly negatively twisted magnetic field lines with $|T_{wn}|\ge 1.5$. In this study, the evolution of these highly negatively twisted field lines is regarded as representing the evolution of the MFR. Around $t=17$, the MFR is gradually formed through magnetic reconnection, and the magnetic flux increases abruptly around 
$t=21$, as indicated by the red circle. This behavior indicates a sudden enhancement of magnetic flux associated with the eruption, implying that magnetic reconnection intensifies dramatically during the initiation of the eruption.

Next we define the unsigned vertical current associated with the MFR at the photospheric boundary as, 
\begin{equation}
I_{MFR}=\int_{|T_{wn}|\ge1.5}|J_z| dS,
\end{equation} 
which represents the integrated current density associated with negatively twisted field lines satisfying $|T_{wn}|\ge 1.5$. Note that $I_{MFR}$ is not identical to the toroidal current ($I_{tor}$) by definition. However, assuming that the MFR forms an isolated current channel, the divergence-free condition of the electric current, ($\nabla\cdot\ {\bf J} =0$), implies that the current flowing through a photospheric footpoint is conserved along the MFR. Therefore, $I_{MFR}$ can be regarded as a proxy for the magnitude of the toroidal current at the apex of the MFR, ($|I_{tor}^{{apex}}|$). Figure \ref{f7}(d) shows the temporal evolution of $I_{MFR}$. First, $I_{MFR}$ exhibits a temporary enhancement around $t=14$, resulting from the interaction between the small-scale emerging flux and the pre-existing coronal magnetic field. Subsequently, similar to the behavior of $\Phi_{|T_{wn}|\ge 1.5}$, the integrated vertical current $I_{MFR}$ increases around $t=21$ in association with the formation of the MFR, followed by a rapid increase during the initiation phase. The vertical current then begins to decrease around
$t=23$.

These results suggest that the initiation of the eruption is primarily governed by a reconnection-driven process rather than by the torus instability. Although the energy may appear to increase, it should be noted that magnetic flux occupied by the field lines with $|T_{wn}|\le 1.5$ decreases, and the total magnetic energy decreases overall. After $t \approx 23$, the vertical current decreases monotonically while $\Phi_{MFR}$ continues to increase, implying that the dominant driving mechanism transitions from the reconnection-driven stage to the hoop force-driven stage. 
 
\subsection{Dependency of Parameters in Small-scale Emerging Flux}
In this section, we present the results for different parameter sets of the small-scale emerging flux. In Runs 2 and 3 (see Table 1), the radius of the small-scale emerging flux is increased, while the magnitude of the normal component of the magnetic field is decreased so as to keep the total magnetic flux constant. In Runs 4 and 5, the magnitude of the normal component of the magnetic field is increased, resulting in an increased flux ratio of the small-scale emerging flux to the total flux (see also Table 1).

\begin{figure}[h]
    \centering
    \includegraphics[width=0.8\textwidth]{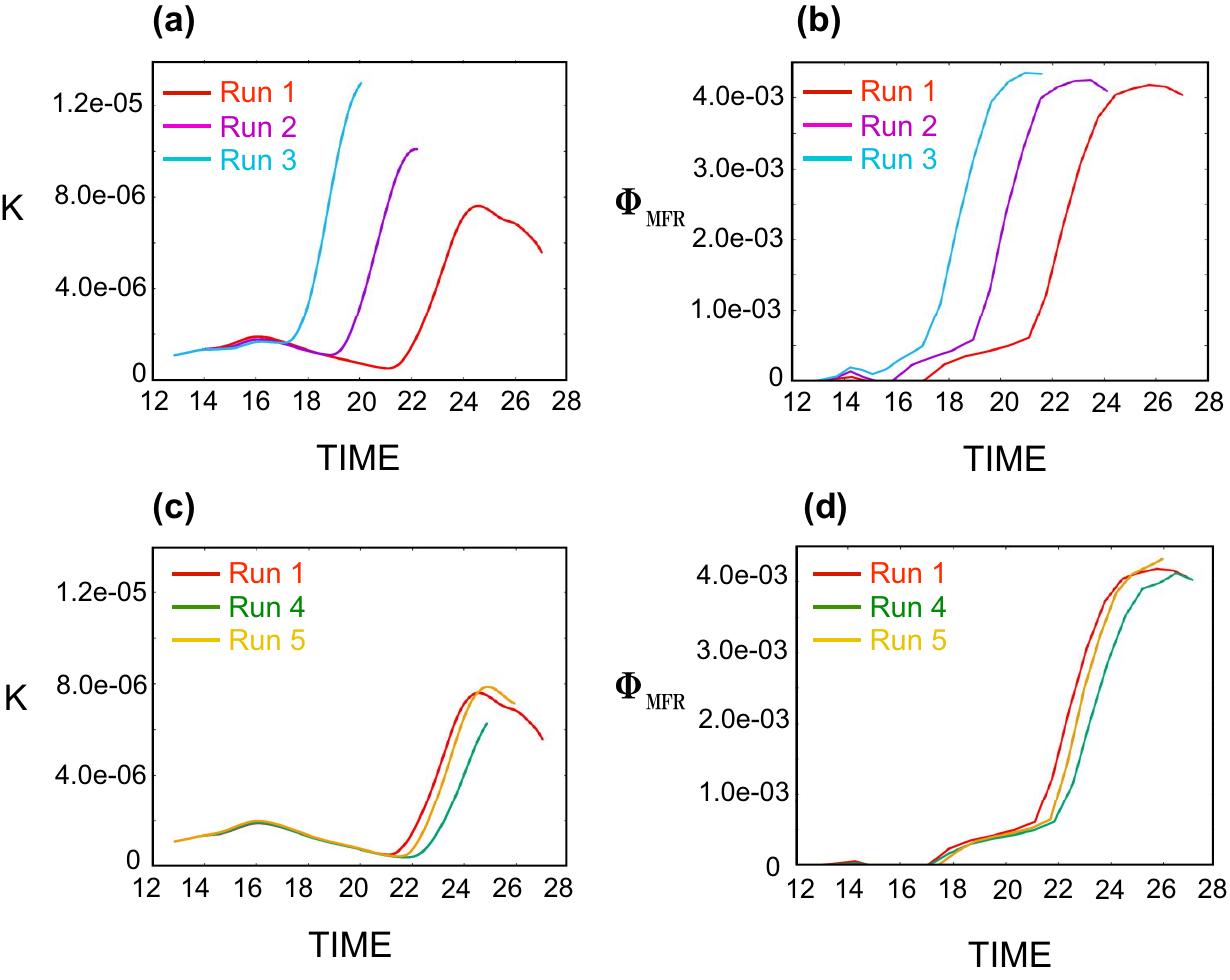}
    \caption{ (a) Temporal evolution of the kinetic energy for Runs 1–3 (red, purple, and light blue, respectively). (b) Temporal evolution of the magnetic flux of the MFR $\Phi_{MFR}$, defined as the region with highly twisted field lines satisfying $|T_w| \ge 1.5$, for Runs 1–3. (c)–(d) Temporal evolution of the kinetic energy and the magnetic flux of the MFR for Runs 1, 4, and 5 (red, Green, and yellow. respectively).}
    \label{f8}
\end{figure}

Figure \ref{f8}(a) shows the temporal evolution of the kinetic energy for Runs 1–3, plotted in red, purple, and light blue, respectively. Although the total magnetic flux of the small-scale emerging region is kept constant, the eruption onset occurs earlier as the size of the emerging flux increases. Figure \ref{f8}(b) presents the temporal evolution of the magnetic flux contained in the MFR for each run. During the early phase of the evolution, the MFR acquires a larger magnetic flux when a larger emerging region is imposed. However, in the later phase, the magnetic flux of the MFR converges to nearly the same value in all runs.

These results suggest that the involvement of highly twisted field lines, {\it i.e}, a formation of highly twisted MFR during the early pre-eruption phase plays an important role in determining the eruption dynamics. Figure \ref{f1}(e) shows the three-dimensional magnetic field configuration before the eruption: the shorter field lines (orange) cross the central PIL, while the longer field lines with relatively large twist (light blue) are located on both sides of the shorter ones. Therefore, a larger small-scale emerging flux can incorporate a greater number of highly twisted field lines. The formation of an MFR composed of highly twisted field lines is likely important in determining the onset time of the eruption.

Figures \ref{f8}(c) and (d) show the temporal evolution of the kinetic energy and the magnetic flux of the MFR for Run 1 and Runs 4–5, respectively. Although the total magnetic flux of the emerging region is increased while keeping its size fixed, the overall evolutionary profiles are very similar, with only slight differences among the runs. Thus, increasing the total magnetic flux of the emerging region is not efficient when it is imposed in an area dominated by less twisted field lines.

 \section{Discussion} \label{sec:dis}
\subsection{An Effect of Small-scale emerging flux Prior to Eruption}
\begin{figure}[h]
    \centering
    \includegraphics[width=0.7\textwidth]{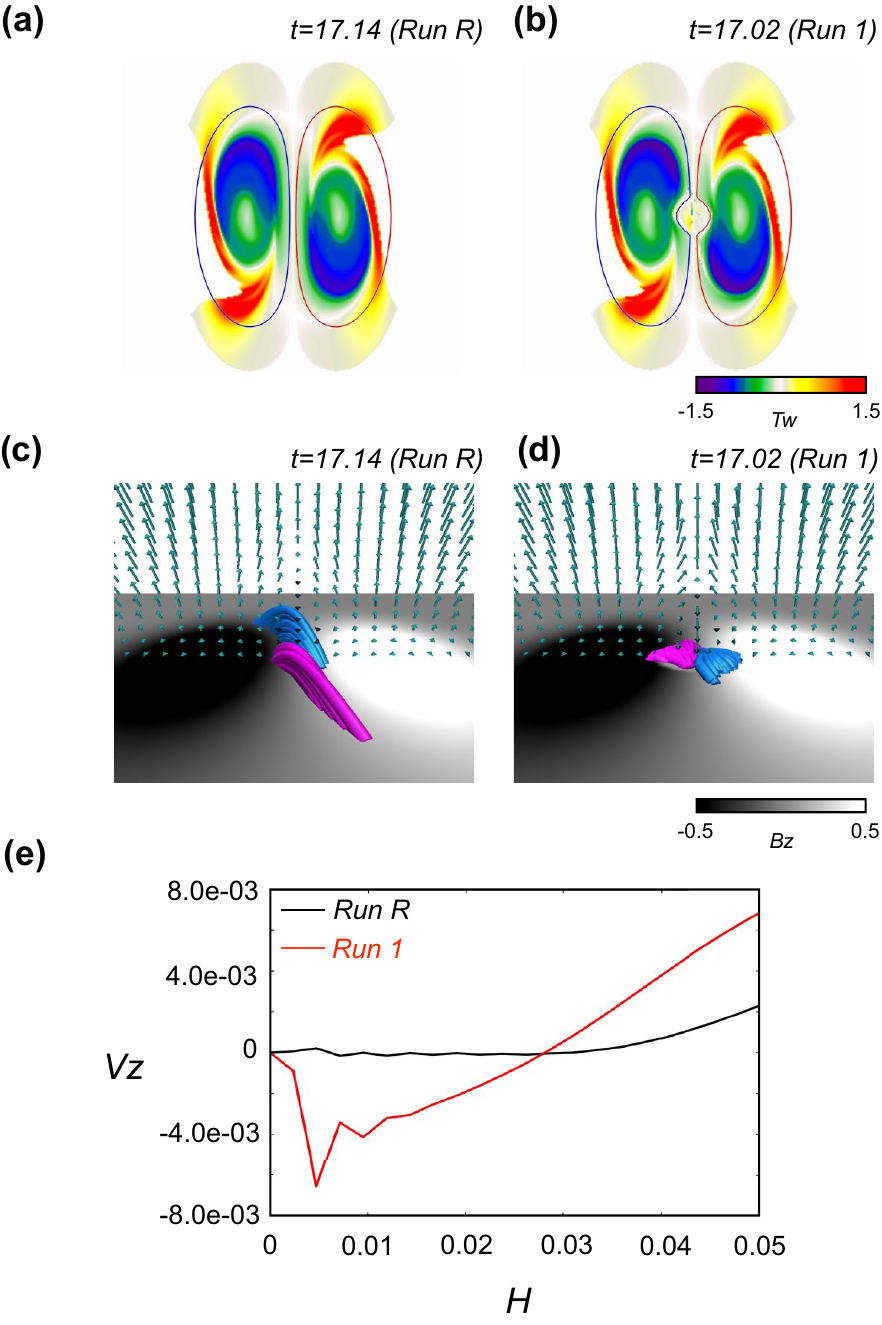}
    \caption{(a) Twist distribution at $t=17.14$ and $t=17.02$ in Runs R and 1, respectively, where the twist of each field lines is mapped onto the bottom surface. The color coding is the same as in the figures above. (b)Three-dimensional magnetic field line structure in the local area where the small-scale emerging flux appears in Run 1. Purple and blue field lines are traced from the positive and negative polarities of the small-scale emerging flux, respectively. Arrows indicate the velocity field defined by $V_x$ and $V_z$. (c) Height profile of the vertical velocity $V_z$ at the center of the small-scale emerging flux, ${\bf r}=(0.5,0.5)$. The black and red curves indicate the vertical velocity profiles in Runs R and 1, respectively.}
    \label{f9}
\end{figure}

We discuss the effect of small-scale emerging flux prior to the eruption. Figures~\ref{f9}(a) and (b) show the twist distribution mapped onto the bottom surface prior to the eruption in Runs~R and~1, respectively. In Run~1, the central circular region corresponds to the small-scale emerging flux, where the twist appears to be relaxed. Figures~\ref{f9}(c) and (d) present three-dimensional views of the magnetic field lines together with velocity arrows composed of $v_x$ and $v_z$ in Runs~R and~1, respectively. In Run~1, compared with Run~R, the sheared field lines on the PIL are removed and a strong downward velocity is generated. Figure~\ref{f9}(e) shows the 1D vertical velocity distribution at the center of the numerical domain in Runs~R and~1, respectively, where this location corresponds to the center of the small-scale emerging flux in Run~1. A clear downward velocity is observed in Run~1, whereas no downward velocity is detected in Run~R.

This result suggests that the small-scale emerging flux removes magnetic pressure from above, resulting in enhanced reconnection, thinning of the current sheet, and the occurrence of rapid reconnection. Therefore, the onset of the eruption can occur earlier than in the case without small-scale emerging flux. This process is consistent with eruption mechanism proposed by \cite{Chen_2000} and \cite{Kusano2004}.
\citealt{Inoue2025} also proposed that small-scale emerging flux enhances tether-cutting reconnection above the polarity inversion line, resulting in accelerating torus instability in the subsequent main phase. While the situation differs, in both cases small-scale emerging flux plays a similar role in promoting magnetic reconnection occurring above and accelerates the destabilization of the coronal magnetic configuration.

\subsection{Potential for Flare Triggering}
Finally, we discuss the potential of small-scale emerging flux to trigger flares. In our simulations, an eruption does not occur immediately after the small-scale emerging flux is imposed. However, as shown in the previous sections, our results suggest that the formation of a highly twisted MFR during the pre-eruption stage plays a crucial role in determining the eruption onset time. This process strongly depends on the initial magnetic configuration.

In the present study, shorter and less twisted field lines cross the PIL interact with the small-scale emerging flux. This interaction may split the less twisted field lines into two parts (\citealt{Kusano2004} and also see Figure 5(b) in \citealt{Inoue2025}), reducing the magnetic pressure as discussed above, and subsequently trigger magnetic reconnection between the longer, highly twisted field lines that are adjacent to the less twisted ones. In this process, it takes time until the longer twisted field lines begin to reconnect. In contrast, $\delta$-type sunspots are well known to produce major flares (\citealt{Sammis2000,Toriumi2019}), and in several cases longer and highly twisted field lines are along the PIL ({\it e.g.}, \citealt{Inoue2014b}). In such a configuration, small-scale emerging flux can readily interact with highly twisted field lines and form a longer, more highly twisted MFR in a region of strong current density, from which magnetic flux is efficiently removed. As a result, this process may lead to faster magnetic reconnection (\citealt{Chaowei2021}) and potentially trigger an eruption shortly after the small-scale flux emerges.

As future work, improvements to the initial magnetic configuration are required to confirm above scenario. If an active region in which highly twisted field lines are located adjacent to the PIL is adopted as the initial condition, how would the eruption process differ from that obtained in the present study? It remains to be clarified whether small-scale emerging flux can act as a direct trigger of solar flares.

\section{Summary} \label{sec:sum}
We performed MHD simulations of solar eruptions to investigate the effect of small-scale emerging flux appearing prior to the onset. Although several previous studies have investigated the role of small-scale emerging flux in solar eruptions, the present study is particularly similar to \cite{Kusano2012} and \cite{Inoue2025} in that no MFR exists in the initial condition. \cite{Kusano2012} focused on the local region around the PIL, while the entire computational domain was described by a linear force-free field. In contrast, \cite{Inoue2025} conducted MHD simulations in a realistic magnetic environment of AR 12371. However, their initial magnetic field was not in equilibrium in the presence of finite resistivity, and the observed small-scale emerging flux was imposed after the initiation of the eruption, i.e., after the onset of tether-cutting reconnection. In the present study, although we employ a simplified magnetic configuration, the computational domain covers the entire active region consisting of both sheared and less-sheared field lines. Furthermore, small-scale emerging flux with a size and magnetic flux comparable to those observed is imposed before the initiation of the eruption. As a result, we find that small-scale emerging flux can lead to an earlier eruption compared with the case without it. The small-scale emerging flux removes magnetic pressure from above, resulting in thinning of the current layer and, eventually, the occurrence of fast reconnection in the thin current sheet. Therefore, in the present study, the small-scale emerging flux does not directly trigger the eruption immediately. Instead, it promotes the formation of a thinner current sheet and facilitates the destabilization of the coronal magnetic configuration, which eventually loses stability through fast magnetic reconnection. The eruption mechanism found in this study is similar to that of the RS (reverse-shear) model proposed by \cite{Kusano2012}. However, in the present study, the polarity of the small-scale emerging flux corresponds to the OP (opposite-polarity) configuration rather than the RS-type configuration. Therefore, depending on the surrounding magnetic configuration, different eruption processes can occur. Furthermore, the enhancement of magnetic reconnection found in the present study is similar to that reported by \cite{Inoue2025}, suggesting that the underlying physical process identified in these studies is common.


In this study, magnetic reconnection gradually develops after the twisting motion is halted for numerical reasons, and we find that small-scale emerging flux simply enhances this reconnection. As a result, the erupting MFR, defined as a cluster of highly twisted field lines with $|T_w|\ge 1.5$ , is created during the pre-eruption stage. Nevertheless, we found that the reconnection-driven process dominates over the hoop force-driven process. 
This result suggests that, once a strong current layer is already formed prior to the eruption and reconnection begins there, the timescale of reconnection is shorter than that of the growth of the torus instability.


Unfortunately, the modeled small-scale emerging flux in this study does not fully capture its observational counterparts. For example, its size is larger than observed values reported by \citealt{Wang_2017}, mainly due to the limitations of numerical resolution. In future work, higher-resolution simulations will be required to model smaller-scale emerging flux and to adopt an initial condition in which longer sheared field lines lie closer to the PIL, in order to clarify how such small-scale flux affects the eruptive dynamics.  Furthermore, many small-scale emerging flux concentrations exhibit complex structures or migrate along the PIL \citep{Kubo2007,Matsumoto2025b}. Such motions along the PIL may play an important role in solar eruptions. \cite{LiuN2025, LiuN2026} pointed out that local magnetic reconnection in active regions is crucial in determining the flare-triggering process and the subsequent dynamics. Therefore, understanding where small-scale magnetic flux emerges and how it evolves or migrates near the PIL is essential for clarifying the flare-triggering mechanism.

Recent high-resolution observations by \citealt{Okamoto2018} reported local magnetic fields as strong as 6000 $G$. It remains an open question what would occur if such strong magnetic fields were to appear locally within an active region and to be relevance to solar eruptions. Therefore, not only simulation studies but also coordinated observations with high-resolution facilities such as BBSO (\citealt{Goode2010,Wenda2010}) and DKIST (\citealt{Rimmele2020}) are more important in future work.

 We thank the referee for valuable comments. This work was supported by National Science Foundation, AGS-2145253, 2309939, and National Aeronautics and Space Administration under grants, 80NSSC23K0406, 80NSSC24M0174. We are grateful to Undergraduate Research and Innovation (URI) Program at the New Jersey Institute of Technology (NJIT) for supporting our summer research. The visualization of magnetic field lines and the presentation of parameters were produced using VAPOR, a product of the National Center for Atmospheric Research (NCAR, cite). All calculations of the MHD simulations in this paper were performed using the computing facilities of the High Performance Computing Center at NJIT.

\bibliography{ref}
\bibliographystyle{aasjournal}

\end{document}